\documentclass[%
 journal=ancac3,
 manuscript=article,
 etalmode=truncate,
 maxauthors=20
 ]{achemso}
\usepackage{pdfpages}

\usepackage{amsmath}
\usepackage{graphicx}
\usepackage{dcolumn}
\usepackage{bm}
\usepackage{mathtools}
\usepackage{amssymb}
\usepackage{csvsimple}
\usepackage[capitalize]{cleveref}
\usepackage[per-mode=reciprocal]{siunitx}
\usepackage{braket}
\usepackage{float}
\usepackage[version=3]{mhchem} 

\author{S. Alex Breitweiser}
\affiliation{These two authors contributed equally}
\alsoaffiliation{Quantum Engineering Laboratory, Department of Electrical and Systems Engineering, University of Pennsylvania, 200 S. 33rd St. Philadelphia, Pennsylvania, 19104, USA}
\alsoaffiliation{Department of Physics and Astronomy, University of Pennsylvania, 209 S. 33rd St. Philadelphia, Pennsylvania 19104, USA}
\altaffiliation{Present address: IBM Research, 1101 Kitchawan Rd Yorktown Heights, NY 10598, USA}

\author{Mathieu Ouellet}
\affiliation{These two authors contributed equally}
\alsoaffiliation{Quantum Engineering Laboratory, Department of Electrical and Systems Engineering, University of Pennsylvania, 200 S. 33rd St. Philadelphia, Pennsylvania, 19104, USA}

\author{Tzu-Yung Huang}
\affiliation{Quantum Engineering Laboratory, Department of Electrical and Systems Engineering, University of Pennsylvania, 200 S. 33rd St. Philadelphia, Pennsylvania, 19104, USA}
\altaffiliation{Present address: Nokia Bell Labs, 600 Mountain Avenue, Murray Hill, NJ 07974, USA}

\author{Tim H. Taminiau}
\affiliation{ 
QuTech, Delft University of Technology, PO Box 5046, 2600, GA Delft, The Netherlands
}
\alsoaffiliation{ 
Kavli Institute of Nanoscience Delft, Delft University of Technology, PO Box 5046, 2600, GA Delft, The Netherlands
}

\author{Lee C. Bassett}
\email{lbassett@seas.upenn.edu}
\affiliation{Quantum Engineering Laboratory, Department of Electrical and Systems Engineering, University of Pennsylvania, 200 S. 33rd St. Philadelphia, Pennsylvania, 19104, USA}

\title[]
  {Quadrupolar resonance spectroscopy of individual nuclei using a room-temperature quantum sensor}


\begin{document}


\begin{abstract}
Nuclear quadrupolar resonance (NQR) spectroscopy reveals chemical bonding patterns in materials and molecules through the unique coupling between nuclear spins and local fields.
However, traditional NQR techniques require macroscopic ensembles of nuclei to yield a detectable signal, which obscures molecule-to-molecule variations.
Solid-state spin qubits, such as the nitrogen-vacancy (NV) center in diamond, facilitate the detection and control of individual nuclei through their local magnetic couplings.
Here, we use NV centers to perform NQR spectroscopy on their associated nitrogen-14 ($^{14}$N) nuclei at room temperature. 
In mapping the nuclear quadrupolar Hamiltonian, we resolve minute variations between individual nuclei.
The measurements reveal correlations between the Hamiltonian parameters associated with the NV center's electronic and nuclear spin states, as well as a previously unreported symmetry-breaking quadrupolar term. 
We further design pulse sequences to initialize, readout, and control the quantum evolution of the $^{14}$N nuclear state using the nuclear quadrupolar Hamiltonian.
\end{abstract}

\vspace{15mm}
Nuclear Quadrupole Resonance (NQR) spectroscopy detects interactions between nuclear electric quadrupole moments and local electric field gradients, aiding the study of molecular structures at low bias magnetic field  \cite{das1958nuclear,zax1985zero,silani2023nuclear}.
NQR spectroscopy is widely applied in security for explosive and drug detection \cite{Grechishkin1997,kim2014polarization,malone20201h}, pharmaceutical analysis of powders \cite{Balchin2005,trontelj2020nuclear,latosinnska2007applications}, and thermometry \cite{vanier1965nuclear,huebner1999nqr}.
Due to the unique fields experienced by nuclei at each site, set primarily by the valence electrons and, therefore, the corresponding chemical bonds, NQR studies reveal a wealth of information that can be used to identify and characterize molecules and bulk materials.
However, due to the small magnetic signal generated by each nucleus, traditional radio-frequency NQR is limited to use with macroscopic samples that contain large nuclear ensembles.
The ability to perform NQR on individual nuclear sites would open up the possibility of studying molecule-to-molecule variations and dynamical changes due to local fields and structural changes, \textit{e.g.},protein folding and drug-target interactions.

Quantum sensors based on optically active defects in semiconductors allow for investigations of much smaller nuclear ensembles.
Defect-based quantum sensors such as the diamond nitrogen-vacancy (NV) center host electronic spin states that can be initialized and measured with laser light and manipulated with microwave signals at room temperature.
The electron spin qubits interact with proximal nuclear spins through unique magnetic hyperfine couplings that are determined by their positions \cite{childress2006coherent}.
Using Dynamical Decoupling (DD) control sequences, it is possible to resonantly amplify these hyperfine couplings \cite{Taminiau2012}, allowing high precision characterization and control of individual nuclei \cite{Taminiau2014, van2012decoherence,zhao2014dynamical,vorobyov2022addressing,cappellaro2009coherence,lang2015dynamical}.
NV-center quantum sensors have been employed along with DD sequences to perform NQR spectroscopy of small nuclear ensembles in deuterated molecules \cite{Lovchinsky2016} and in hexagonal boron nitride crystals \cite{Lovchinsky2017,Henshaw2022}.
In other regimes, NV-center ensembles have been used to boost the sensitivity of traditional NQR detectors for macroscopic powder samples \cite{silani2023nuclear}.
However, accessing individual nuclei and retrieving their quadrupolar Hamiltonian has remained an open challenge.

In this work, we demonstrate DD-based, room-temperature NQR spectroscopy of the nitrogen-14 ($^{14}$N) nuclei intrinsic to individual NV centers.
In this way, the NV centers serve as both quantum sensors and as analogs of individual molecules.
The measurements reveal considerable variations in the $^{14}$N quadrupolar and hyperfine parameters among different NV centers, as well as a previously unreported term in the nuclear quadrupolar Hamiltonian that results from symmetry breaking. 
We further observe correlations between the nuclear Hamiltonian parameters and the electronic Zero-Field Splitting (ZFS) parameters, highlighting the potential of NQR spectroscopy to reveal details of local chemical structure and deformations due to electric or strain fields.
Finally, we design and implement DD sequences that utilize the $^{14}$N quadrupolar Hamiltonian to facilitate initialization and arbitrary quantum control of the $^{14}$N nuclear spin.

\begin{figure}
\includegraphics[]{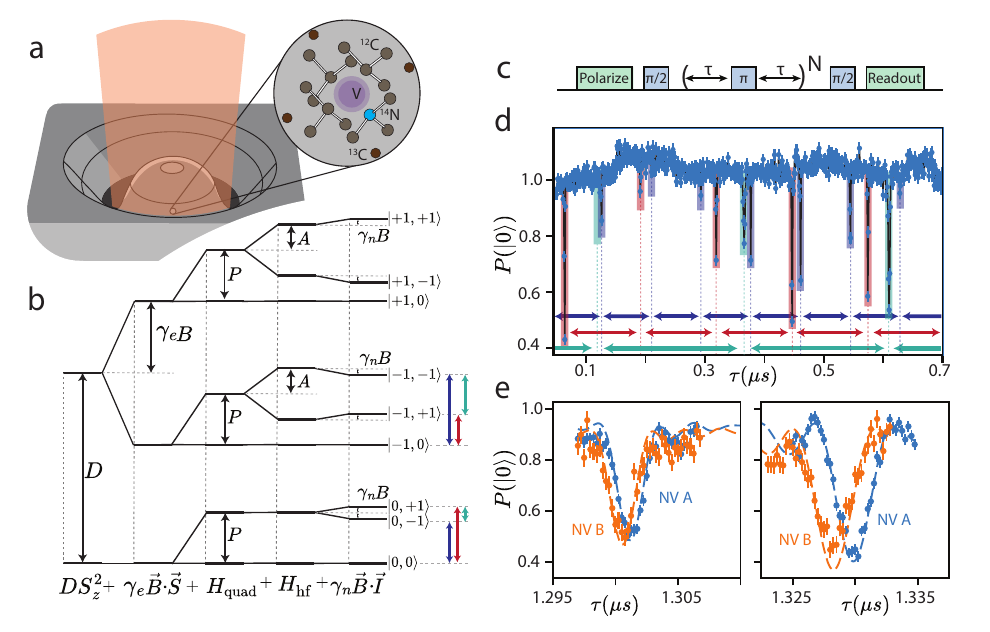}
\caption{\label{Fig1} \textbf{Electron-nuclear interactions in $^{14}$NV centers} (a) Model of a diamond NV center located within a solid immersion lens, composed of an electronic spin (purple) coupled to a $^{14}$N nucleus (blue).
(b)  Pictoral representation of the Hamiltonian terms for $^{14}$NV. 
(c) Schematic of a DD spectroscopy sequence. Initial and final $\frac{\pi}{2}$ pulses are in opposite directions, while decoupling $\pi$ pulses are XY8 symmetrized.
(d) DD NQR features corresponding to the $\ket{m_I=\pm 1}$ to $\ket{m_I=0}$ transitions (dips marked purple and red) appear in the presence of an off-axis magnetic field ( $N=64$, $B=\SI{193}{G}$).
A third series of dips corresponding to the $\ket{m_I=\pm 1}$ transition (marked green) appears due to the quadrupolar asymmetry parameter, even without an off-axis magnetic field.
Panel (b) displays the energy levels associated with the peaks.
(e) DD spectra for two different NV centers (blue and orange), here with $N=32$ pulses repetitions, reflect significant differences in the $^{14}$N quadrupolar Hamiltonian. Data are shown with markers, along with best-fit simulation results as dashed lines. }
\end{figure}

\section{\label{sec:nvnqr} Electron-nuclear interactions in diamond NV centers}

The NV center (Figure \ref{Fig1}a) consists of one substitutional $^{14}$N coupled to a vacancy in the diamond lattice.
In its negatively charged state, the NV center hosts an electronic spin-1 state that undergoes a spin-dependent optical pumping transition, allowing the spin state to be initialized and read out optically \cite{doherty2013nitrogen}.
This electronic spin interacts with the intrinsic $^{14}$N nuclear spin within the NV center ($\approx \SI{99.7}{\%}$ spin-1 $^{14}$N in natural abundance), as well as with $^{13}$C nuclei in the surrounding diamond lattice (spin-$\frac{1}{2}$, $\approx \SI{1.1}{\%}$ natural abundance) and any other nearby nuclear spins.
In the isotropic case, ignoring the effects of strain or electric fields, the general Hamiltonian comprising the electron spin interacting with a single nuclear spin is given by 
\begin{equation}\label{eq:FullHamiltonian}
H  = D S_z^2 + E(S_X^2-S_Y^2)+ \gamma_{e} \Vec{B}\cdot \Vec{S} + H_{\text{quad}} + H_{\text{hf}} + \gamma_{n} \Vec{B}\cdot \Vec{I},
\end{equation}
where $\Vec{S} = (S_X,S_Y,S_Z)$ is the electronic spin operator, $\vec{I} = (I_X,I_Y,I_Z)$ is the nuclear spin operator, $\gamma_e$ ($\gamma_n$) is the electronic (nuclear) gyromagnetic ratio, and $B = (B_X,B_Y,B_Z)$ is the external magnetic field. 
Figure \ref{Fig1}b shows this Hamiltonian diagrammatically for the specific example of the NV-center's intrinsic $^{14}$N nucleus. 
The first two terms represent electronic zero-field splitting (ZFS), followed by the Zeeman term for electronic spins. 
This is succeeded by the nuclear quadrupolar term and hyperfine coupling and the Zeeman term for nuclear spins.
The term $H_{\text{hf}}$ represents the hyperfine interaction, which takes the general form
\begin{equation}
H_{\text{hf}} = \vec{S} \cdot \mathbf{A} \cdot \vec{I}
\label{eqn:fullhyp}
\end{equation}
where $\mathbf{A}$ is the hyperfine interaction tensor. 
For $^{14}$NV, $H_{\text{hf}}$ takes the simplified form:
\begin{equation}\label{eq:Hhf_sec}
H_{\mathrm{hf}} = A_{Z} S_Z I_Z + A_{\bot}(S_XI_X + S_YI_Y),
\end{equation}
where $A_{Z}$ and $A_{\bot}$ are the parallel and perpendicular hyperfine coupling strengths.
The second term in Eq.~\ref{eq:Hhf_sec} generally does not affect the electron-nuclear dynamics due to the large mismatch in energy splitting between the electron and nuclear spin states, leaving the parallel term as the primary hyperfine-coupling effect.

The term $H_{\text{quad}}$ represents the nuclear quadrupolar Hamiltonian, which is nonzero for nuclear species with total nuclear spin $I\geq 1$.
The quadrupolar Hamiltonian can, in general, be written as \cite{Smith1971}
\begin{equation}
\label{eqn:Hquad_full}
H_\mathrm{quad} = \frac{eQV_{ZZ}}{4I(2I-1)}[3I_Z^2-I(I+1)+\frac{V_{XX}-V_{YY}}{2 V_{ZZ}}(I_+^2 + I_-^2)]
\end{equation}
where $e$ is the electron charge, $Q$ is the quadrupolar moment unique to each nuclear isotope, $V_{ZZ} = \partial^2 V / \partial z^2 $ is the electric field gradient along the principal nuclear axis, and $V_{XX}$ and $V_{YY}$ are the electric field gradients in the perpendicular plane. 
The principal axes are chosen so that $|V_{ZZ}| > |V_{XX}| > |V_{YY}|$, and the electric field gradient is diagonal in this basis.
For a particular nucleus, Eq.~\eqref{eqn:Hquad_full} takes the simplified form
\begin{equation}
H_{\mathrm{quad}} = PI_Z^2 + \alpha (I_+^2 + I_-^2),
\label{eqn:Hquad}
\end{equation}
where $P$ and $\alpha$ are constant parameters representing the quadrupolar splitting and asymmetry parameters, respectively.

Although the nuclear quadrupolar Hamiltonian term is distinct from the electronic spin, its effects on the electronic spin can be observed \textit{via} the hyperfine interaction using DD sequences as shown in Figure \ref{Fig1}c.
Transverse terms in $H_\mathrm{hf}$ (the second term in Eq. \ref{eq:Hhf_sec}) lead to rotations of the nuclear state that depend on the electron spin projection.
DD sequences amplify this interaction since multiple small rotations accumulate when the spacing between pulses is resonant with the hyperfine-shifted frequency of the nuclear Larmor precession, causing resonant series to emerge in DD spectra \cite{Taminiau2012}.

Figure \ref{Fig1}d shows an example of a DD NQR spectrum in which three distinct resonance series can be observed.
The series correspond to electron-spin-dependent transitions between the $^{14}$N nuclear states as indicated in Fig.~\ref{Fig1}b.
As discussed in the next section, the physics responsible for transitions between nuclear states with $\ket{m_I=0} \leftrightarrow \ket{m_I=\pm 1}$ is different from those transitions between $\ket{m_I=\pm1}$ states. 
Nevertheless, the observation of all three resonance series constitutes a complete measurement of the nuclear quadrupolar Hamiltonian (Eq. \ref{eqn:Hquad}), together with $A_Z$.
Since the DD sequence extends the coherence lifetime of the electronic spin, this method allows extremely precise determination of $P$ for each nucleus and also reveals the existence of small $\alpha$ Hamiltonian terms that had previously not been detected nor considered \cite{doherty2013nitrogen}.

\section{\label{sec:results}DD NQR spectroscopy}

The $^{14}$N nucleus intrinsic to the NV center represents a convenient testbed to illustrate the physics of DD-based NQR.
Ideally, the $C_{3v}$ symmetry of the NV center should cause the asymmetry quadrupolar parameter $\alpha$ to vanish. 
Moreover, in the presence of a purely longitudinal magnetic field ($B_X=B_Y=0$), the axially-symmetric hyperfine Hamiltonian of Eq.~\ref{eq:Hhf_sec} does not generate nuclear spin rotations under DD sequences, since the magnetic field direction experienced by the nucleus is independent of the electron spin projection.
In real systems with reduced symmetry, however, both of these conditions are relaxed.

In the presence of a weak transverse magnetic field ($B_X \ll B_Z$) an effective perpendicular hyperfine coupling term appears due to spin mixing \cite{shin2014optically}, leading to an approximate hyperfine Hamiltonian given by
\begin{equation}
H_{\text{hf}} \approx A_Z S_Z I_Z + F\frac{\gamma_{N}B_XA_{\bot}}{\gamma_eB_Z}S_ZI_X,
\end{equation}
where $F$ is a constant that is particular to the $^{14}$N nuclear isotope \cite{Liu2019}.
Previous authors have used this effective hyperfine interaction to observe nuclear quadrupolar interactions for NV ensembles using electron spin-echo envelope modulation \cite{Shin2014}, to perform dc vector magnetometry using single NV centers \cite{Liu2019} and to realize high fidelity gates \cite{bartling2024universal}.
We use it in order to quantify the quadrupolar Hamiltonian parameters \textit{via} DD NQR spectroscopy.
In a DD control sequence with appropriate pulse spacing, the $S_ZI_X$ term in the effective hyperfine Hamiltonian facilitates electron-spin-dependent rotations of the $^{14}$N spin (see Figure \ref{Fig1}c), in analogy with the case for ${13}$C nuclei \cite{Taminiau2012}. 
The rotations manifest in DD spectra as two distinct resonance series, each corresponding to one of the $\ket{m_I} = 0$ to $\ket{m_I = \pm 1}$ transitions, with spacing given by
\begin{equation}
     \tau_k \approx  \frac{(2k+1)\pi}{2P\pm A_z \mp \omega_N }
\end{equation}
where $\omega_N = \gamma_N B $. 

Figure \ref{Fig1}e shows the DD NQR spectra obtained by sweeping the pulse spacing near two such transition resonances for two different NV centers.
The shift in resonance position reflects differences in $P$ and $A_Z$ for these two NV centers.
We use numerical simulations to fit DD NQR spectra acquired using different $N$ around these two resonances; see the Supporting Information, Sec.~II.
Table \ref{Table1} shows NQR spectroscopy results for six NV centers located within the same diamond sample (see the Supporting Information, Fig.~S3, for the underlying data).
Interestingly, the values of $P$ and $A_Z$ show a variance one order of magnitude larger than the measurement uncertainty.
In particular, NV A exhibits values for $P$ and $A_Z$ that differ by several \SI{}{kHz} from the other NVs in this sample.
NV A is also the only NV under a diamond solid immersion lens (SIL).
These milled structures are used to minimize optical losses caused by total internal reflection and spherical aberration (See Figure \ref{Fig1}a) \cite{jamali2014microscopic,wildanger2012solid}.
Yet, they are recognized to influence the local strain field at the NV centers \cite{knauer2020situ}, consequently affecting the NV Hamiltonian\cite{meesala2018strain,assumpcao2023deterministic}.


\begin{table}[h]
\begin{center}
\begin{tabular}{c|c|c|c|c|c}

\textbf{NV} & \textbf{D} $\SI{}{(MHz)}$ & \textbf{E} $\SI{}{(MHz)}$ & $\mathbf{A_Z}$ $\SI{}{(kHz)}$ & \textbf{P} $\SI{}{(kHz)}$ & $\mathbf{\beta}$ \\
\hline 
\hline
A & $\SI{2859.2(0.02)}{}$ & $\SI{8.33(0.04)}{}$ & $\SI{2168.1(0.1)}{}$ & $\SI{4934.9(0.1)}{}$ &  $\SI{0.0016(0.0002)}{}$\\
\hline
B & $\SI{2870.47(0.02)}{}$ & $\SI{7.51(0.04)}{}$ & $\SI{2164.7(0.1)}{}$ & $\SI{4939.5(0.1)}{}$ &   $\SI{0.0039(0.0003)}{}$\\
\hline
C & $\SI{2870.39(0.02)}{}$ & $\SI{7.63(0.04)}{}$ & $\SI{2163.5(0.5)}{}$ & $\SI{4939.4(0.2)}{}$ &   $\SI{0.0095(0.0004)}{}$\\
\hline
D & $\SI{2870.37(0.01)}{}$ & $\SI{7.58(0.03)}{}$ & $\SI{2165.0(0.3)}{}$ & $\SI{4939.2(0.1)}{}$ &   $\SI{0.0175(0.0006)}{}$\\
\hline
E & $\SI{2872.2(0.03)}{}$ & $\SI{7.58(0.04)}{}$ & $\SI{2162.9(0.4)}{}$ & $\SI{4936.9(0.2)}{}$ &  $\SI{0.0205(0.0004)}{}$\\
\hline
F & $\SI{2870.41(0.02)}{}$ & $\SI{7.04(0.03)}{}$ & $\SI{2162.9(0.4)}{}$ & $\SI{4940.7(0.2)}{}$ &   $\SI{0.0082(0.0004)}{}$\\
\end{tabular}
\end{center}
\caption[]{\label{Table1} Electronic ZFS, hyperfine, and $^{14}$N quadrupolar parameters for each NV studied.}

\end{table}

\section{Forbidden quadrupolar transitions}

\begin{figure}
\includegraphics[]{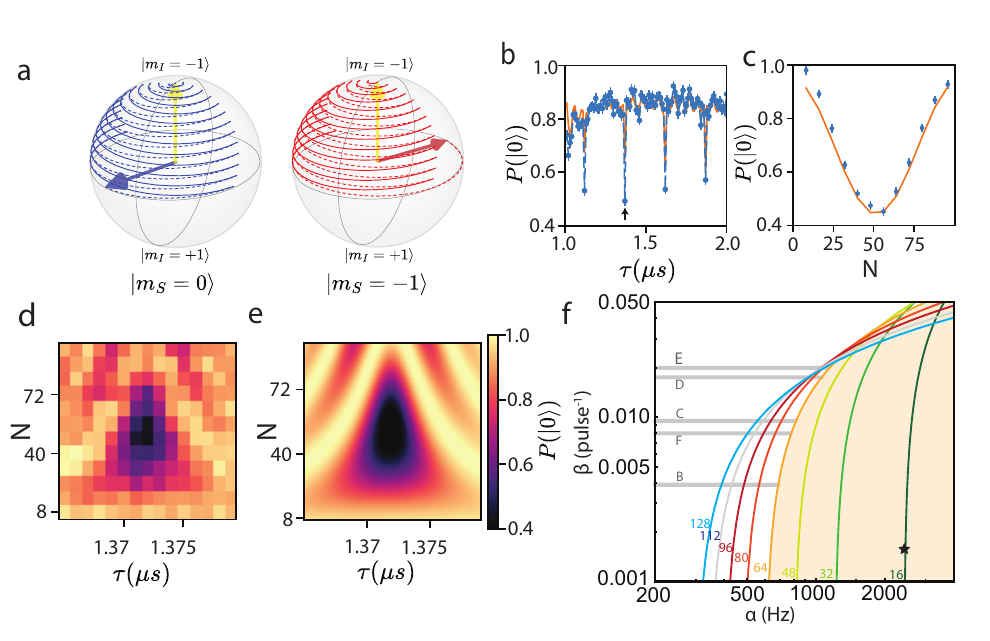}
\caption{\label{Fig3} \textbf{Forbidden transitions} (a) Evolution of the nuclear spin in the $\ket{m_I=\pm1}$ manifold when the electronic spin is in the state $\ket{m_s=0}$ (blue) and $\ket{m_s=-1}$ (red) for a tuned DD sequence with $N=12$.
(b) DD spectroscopy data for NV A (blue points with dashed line) and simulation (solid orange curve) for $N=32$.
(c) The measured electron spin projection (blue markers) as a function of $N$ for fixed $\tau = \SI{1.372}{\mu s}$ (black marker in (b)) agrees with simulations (solid orange curve). 
(d) DD spectroscopy data as a function of $\tau$ and $N$, which is fit using simulations (e) to determine the value of $\alpha$. Error bars in (d) are comparable to those in (b) and (c).
(f) Detection limit for $\alpha$ as a function of $\beta$ for tuned DD sequences with different $N$; the detection limit is the situation where the signal-to-noise ratio exceeds 1.
The noise floor is assumed to be constant at $P(\ket{0})=0.2$.
The shaded area represents the detection threshold for the sequence applied experimentally. 
The star marks the value for NV A.
} 
\end{figure}

Although the $\alpha_i$ term in the $^{14}$N quadrupolar Hamiltonian vanishes for the ideal case of $C_{3v}$ symmetry, local perturbations such as strain and electric fields can distort the electronic wavefunctions, leading to nonzero transverse electric-field gradients at the $^{14}$N position.
When $\alpha$ is nonzero, the second term in Eq.~\eqref{eqn:Hquad} directly couples the $\ket{m_I=-1}$ and $\ket{m_I=+1}$ nuclear states, causing nuclear transitions to occur that are typically symmetry forbidden.
Figure \ref{Fig3}a illustrates these dynamics for a suitably tuned DD sequence, whereby the nuclear spin evolves in the $\ket{m_I=+1}$ manifold according to slightly different rotation axes depending on the electronic spin state; after many pulses, the nuclear spin evolves into orthogonal spin states. 
The resulting entanglement between electronic and nuclear spins manifests as a reduced signal amplitude for these carefully tuned pulse sequences.
DD spectroscopy of NV A (Fig.~\ref{Fig3}b) reveals the presence of these forbidden transitions as a series of sharp, periodic resonances with a spacing given by
\begin{align}
  \tau_k \approx \frac{ (2 k +1)\pi}{ 2 (A_Z-2\omega_ N) },
\end{align}
where $\omega_N = \gamma_N B $. 
Section~VII in the Supporting Information includes a derivation of this expression. 

In analogy to the typical phenomena of DD resonances (as in Fig.~\ref{Fig1}c), whereby $A_\perp$ induces $S_z$-dependent rotations between states with $\Delta m_I=1$, here the nonzero $\alpha$ term induces $S_z$-dependent transitions between $\ket{m_I = \pm1}$ states.
When $\alpha$ is small, many pulses are needed in order to accumulate a measurable rotation angle.
Figure \ref{Fig3}c shows the evolution of the DD signal as a function of $N$; the contrast is reduced by approximately $\frac{1}{3}$ due to the thermal occupation probability of the uncoupled $\ket{m_I=0}$ state.
By varying both $\tau$ and $N$ (Fig.~\ref{Fig3}f) around a particular resonance, we map out the full dynamics of these forbidden quadrupolar resonances.
A fit using numerical simulations (Fig.~\ref{Fig3}f) yields a best-fit value of $\alpha = 2\pi\times\SI{2.429(12)}{kHz}$.
The ratio $\alpha/P=5\times 10^{-4}$ illustrates how a tiny Hamiltonian parameter can have a substantial impact on nuclear dynamics and be measured with high precision using DD spectroscopy. 

The sensitivity of the DD-based measurement is limited by intrinsic decoherence mechanisms (captured by $T_2$) and by pulse errors \cite{ahmed2013robustness}.
For the sequences we consider, the total experimental sequence times are much shorter than the intrinsic decoherence time (typically $T_2\approx 1$ms at room temperature), and the contrast decay is dominated by pulse errors, which we model using an exponential envelope, $e^{-\beta N}$.
This envelope constrains the practical detection limit of $\alpha$, shown in Figure \ref{Fig3}b as a set of curves for different $N$ as a function of $\beta$.

Of the six NV centers we studied, we only observed forbidden transitions for NV A.
We propose two reasons for this observation. 
First, NV A may experience larger-than-average symmetry breaking from transverse strain or electric fields due to its location at the center of a milled SIL, and hence a larger value of $\alpha$.
This is supported by measurements of the electronic ZFS $D$ and $E$ (Table~\ref{Table1}) are significantly different than for other NV centers in the sample.
Moreover, NV A features the smallest $\beta$ of the sample and, subsequently, the lowest detection limit.
Figure~\ref{Fig3}f shows the $\beta$ values for the other NV centers along with a shaded region corresponding to the $N<64$ limit we experimentally investigated; we expect that $\alpha$ for these NV centers is outside the detection region.

\section{Nuclear initialization and coherent evolution}

\begin{figure}
\includegraphics[width=0.975\linewidth]{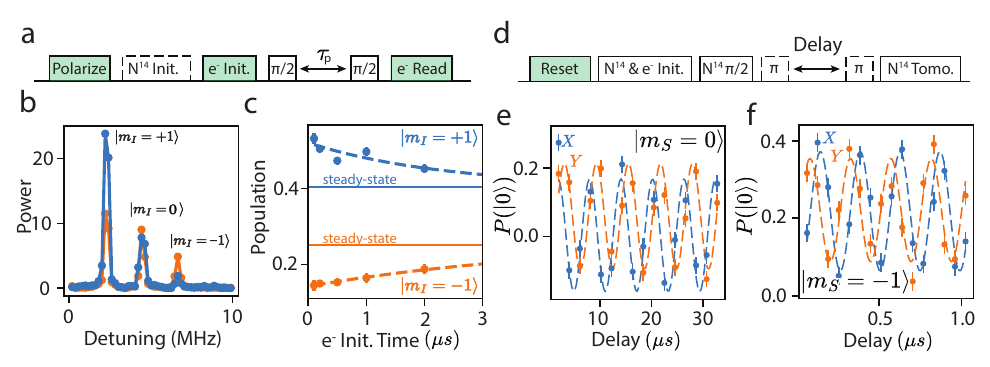}
\caption{\label{Fig4} \textbf{Initialization and Coherent 
Evolution} (a) Experimental sequence used to probe initialization of the $^{14}$N nuclear spin. 
(b) Power spectra of the free-precession data acquired without (orange) and with (blue) a DD-based initialization sequence. 
(c) $\ket{m_I=+1}$ (blue markers) and $\ket{m_I=-1}$ (orange markers) populations fitted from Ramsey data while varying the amount of green time used to reinitialize the electron state. 
Exponential fits (dotted lines) show the population difference decaying to the steady state values (solid horizontal lines).
(d) Pulse sequence used to measure the free induction decay of the $^{14}$N nuclear spin. 
(e,f) Oscillations of the X (blue markers) and Y (orange markers) projection of the $^{14}$N nuclear spin within the $\ket{m_I=+1}/\ket{m_I=-1}$ manifold during free evolution while the electron spin is in the (e) $\ket{m_S=0}$ and (f) $\ket{m_S=-1}$ state.
 } 
\end{figure}

In addition to their use in sensing, DD sequences can be used to achieve precise control over individual spin states \cite{Taminiau2012,van2012decoherence,Taminiau2014}.  
Combinations of conditional and non-conditional gates can be used to construct protocols for nuclear-spin initialization, unitary control, and entanglement with the electron spin \cite{Taminiau2014}.
Figure~\ref{Fig4}a shows a sequence used to probe $^{14}$N spin initialization.
Here, two DD sequences functioning as CNOT gates transfer the population from the electron to the nuclear spin states, and a subsequent electron-spin free-precession sequence probes the resulting nuclear population.
The electron precession exhibits three oscillation frequencies associated with the $^{14}$N spin states, resolved by the $A_Z$ hyperfine coupling (Fig.~\ref{Fig4}b). 
The relative amplitudes of these three oscillations, extracted from the power spectrum (Fig.~\ref{Fig4}b) reflect the nuclear spin occupation probabilities. 
In this case, the DD initialization sequence applied to a forbidden transition of NV A transfers population from the $\ket{m_I=-1}$ state directly to $\ket{m_I=+1}$, further confirming the physical interpretation of these resonances.

The data in Fig~\ref{Fig4}b show that the $^{14}$N nuclear spin is partially polarized even without the  $^{14}$N initialization sequence. 
This is due to the off-axis hyperfine interaction in the optically excited state \cite{busaite2020dynamic}. 
By sweeping the duration of green illumination used to reset the electron spin, the non-equilibrium nuclear population lasts for several microseconds (Figure \ref{Fig4}c) before returning to the steady state values, consistent with other studies on the $^{14}$N nuclear spin population \cite{Chakraborty2017}. 
Nuclear tomography on the  $^{14}$N using the electronic spin confirms the effectiveness of the initialization sequence (Figure \ref{Fig4}d).
Oscillations in the $^{14}$N nuclear spin projection within the $\ket{m_I=+1}/\ket{m_I=-1}$ manifold during free evolution, observable while the electron spin is in the$\ket{m_S=0}$ and $\ket{m_S=-1}$ states, are evident in Figures \ref{Fig4}e and \ref{Fig4}f. 
Section~V of the Supporting Information includes further information about the initialization and tomography sequences.

\begin{figure}
\includegraphics[width=0.95\linewidth]{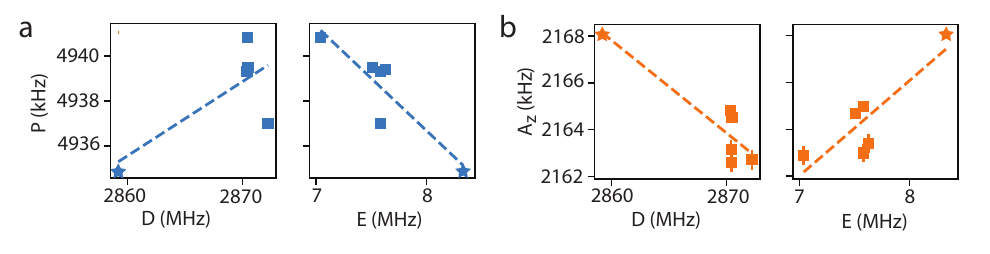}
\caption{\label{Fig5} 
\textbf{Impact of ZFS parameters} (a)
Measured $P$  parameters for $^{14}$N, derived from DD spectroscopy, plotted against the electron ZFS parameters $D$ and $E$ electron splitting parameters derived from zero-field ESR data. 
(b) Measured $A_z$ parameters for $^{14}$N, similarly plotted against $D$ and $E$.
Dashed lines indicate linear fits. 
Error bars are comparable in size to the markers. NV A is indicated in each plot by a star.
 } 
\end{figure}

\section{\label{sec:comparison}Comparisons of NQR and ZFS parameters}

Figure \ref{Fig5} shows the measured values for the $^{14}$N quadrupolar splitting $P$ and hyperfine coupling $A_z$ plotted against the electronic ZFS parameters $D$ and $E$ for each NV center studied.
The clear correlation illustrates the influence of local strain and electric fields on the quadrupolar Hamiltonian.
The measurements cannot be explained using the Hamiltonian in eqn.~(\ref{eq:FullHamiltonian}) alone with fixed quadrupolar parameters (see Supporting Information, Sec.~VII); rather, distortions of the chemical bonds cause changes in the local electric-field gradients that determine $P$ and $\alpha$.
As discussed earlier, NV A exhibits ZFS parameters that are significantly shifted from the mean, consistent with a large local strain or electric field. 
NV A is also the only center for which we observed a nonzero $\alpha$ parameters in the $^{14}$N quadrupolar Hamiltonian.
Using Equation \ref{eqn:Hquad_full}, we obtain $V_{zz} = 1.359(7) \times 10^{22}\, \text{V/m}^2$, and the fitted value of $\alpha$ for NV A gives a normalized transverse electric field gradient value of $\frac{V_{XX}-V_{YY}}{ V_{ZZ}} = \SI{0.0181 (3)}{}$.


\section{Conclusion}

This study introduced a method to measure the quadropolar Hamiltonian of an individual nucleus using a single electronic spin as a sensor.
The method enabled the observation of previously unidentified terms in the  $^{14}$N Hamiltonian for NV centers, and elucidated correlations between the electronic and nuclear Hailtonian parameters due to distortions of the defect's structure. 
Compared to existing techniques, this approach offers numerous advantages.
Due to the frequency selectivity of DD spectroscopy, each nucleus is uniquely resolved by its hyperfine coupling to the electron sensor. 
The measurement is also highly local; its sensitivity decreases rapidly with distance since the hyperfine coupling scales as $\sim 1/d^3$, where $d$ is the sensor-target separation; see the Supporting Information for further details on sensitivity limits.
Recent advances in creating and stabilizing shallow NV centers \cite{neethirajan2023controlled,kawai2019nitrogen}, combined with this approach, can potentially allow nanoscale NQR sensors capable of probing individual nuclei at the single-molecule level.
This method can also be used to probe nuclei associated with surface groups, or to fingerprint defects inside the bulk.
Functionalized nanodiamonds containing NV centers can be suspended in liquid solutions and probed in biochemical environments for \textit{in situ} and \textit{in vivo} chemical sensing applications \cite{shulevitz2023nanodiamond,qin2023situ,holzgrafe2020nanoscale}.

One of the significant advantages of the NV center is its surrounding $^{13}$C  ensemble, which can function as a quantum register \cite{Taminiau2014, Bradley2019, Abobeih2019} and enhance sensing capacity \cite{zaiser2016enhancing}.
This approach preserves the ability to utilize such techniques.
Since the hyperfine and quadrupolar parameters are much stronger than the Zeeman splitting under our experimental conditions, the resonance positions remain stable over a wide range of magnetic field values.
Hence, the magnetic field can be tuned for the convenience of the sample/system under study.

The accurate measurement of the asymmetry of the quadrupolar moment is becoming increasingly crucial for precise control and manipulation of quantum systems \cite{wang2017experimental,gentile2021learning,nie2015quantum}.
Quadrupolar asymmetry plays a role, for example, in semiconductor quantum dots \cite{hackmann2015influence} where it is the source of decoherence, and in nuclear spin squeezing \cite{korkmaz2016nuclear} where it can be used for control.
The ability to detect even the small magnitude of the asymmetry in systems where it is expected to be zero represents a significant advancement.

Similar to other pulsed quantum spectroscopy techniques, the sensitivity of this NQR technique is limited by $T_2$ and pulse errors.
Pulse errors can be minimized by implementing more sophisticated control schemes \cite{arroyo2014room, rong2015experimental}.
Although $T_2$ at room temperature is already close to the $T_1$ limit, it is possible to adapt NMR sensing protocols that surpass the $T_2$ limit for use with NQR \cite{Schwartz2019}. 
Additional techniques such as optimizing NV depth \cite{ devience2015nanoscale}, improving sample preparation \cite{cohen2020confined}, and employing machine learning to compensate for noise \cite{aharon2019nv} will further boost the sensitivity.


\section{\label{sec:methods}Methods}

The sample and experimental setup are described in Refs.~\citenum{hopper2016near} and \citenum{Hopper2020}. Important details are included here.

The sample is a type-IIa electronic-grade synthetic diamond (Element Six), treated with 2-MeV electron irradiation ($10^{14} \, \text{cm}^{-2}$) and annealed at 800$^\circ$ C for 1 hour in forming gas. 
NV A is at the focus of a 6-\textmu m-diameter solid immersion lens (SIL) surrounded by a circular antenna used for microwave control.
Details on SIL fabrication are included in the Supporting Information, Sec.~I.


Using the SIL around NV A, we detect $\sim$0.1 photoluminescence photons per measurement shot. 
For other NVs that were within the antenna's range but not within the SIL's focus, we detect $\sim$0.02 photons per shot, leading to reduced signal-to-noise ratio.
Magnetic fields are supplied by a permanent magnet, which was aligned and calibrated using the $\ket{m_S=0}$ to $\ket{m_S=\pm1}$ spin-resonance transitions of NV A. 
For initializing the spin states, $\SI{20}{\mu s}$-long green laser pulses are used to reset the system, while shorter ($\SI{100}{ns}$) laser pulses are used to reinitialize only the electron spin. 

The experiment timing was controlled by a pair of Arbitrary Waveform Generators (AWGs).
One (AWG520 Tektronix) was triggered to start the experiment and controlled the optical excitations and collection paths, including the AOM used to turn on the green (532nm) laser used for readout and initialization, and the data acquisition system (National Instruments, PCIe-6323).
The AWG520 was also used to trigger another AWG (AWG7102 Tektronix) which was used to control the IQ modulation of a benchtop signal generator (SG384,
Stanford Research Systems), which was fed into a high bandwidth mixer (ZX05-63LH+, Mini-Circuits) to allow fast pulses and a high-isolation switch (ZASWA-2-50DR, Mini-Circuits, allowing ) to prevent on-resonance leakage from decohering the spin, both of which are also controlled by the AWG7102.
Interpolated pulse spacings are used to increase the resolution beyond the hardware limitations \cite{Ajoy2017}.
The output was fed through a USB-controlled microwave attenuator (Rudat 6000-60, Mini-Circuits) and broadband amplifier (ZHL-16W-43-S+, Mini-Circuits) before being delivered into the sample through a custom SMA-connected PCB, which is, in turn, wire-bonded to the antenna traces.


\section{Supporting Information}

Details of sample fabrication, experimental data from ESR and Ramsey experiments, dynamical decoupling spectra showing individual carbon-nuclei resonances, fitted simulations of Hamiltonian parameters, nuclear spin initialization sequences, and analysis of transverse electric field effects on the observed resonances.

\begin{acknowledgement}
This work was primarily supported by the National Science Foundation under awards ECCS-1842655 (S.A.B., T.-Y. H., and L.C.B.) and DMR-2019444 (M.O. and L.C.B.). S.A.B. acknowledges support from an IBM PhD Fellowship. M.O. acknowledges support from the Natural Sciences and Engineering Research Council of Canada (NSERC). We thank Amelia Klein and Joseph Minnella for fruitful discussions and critical reading of our manuscript.
\end{acknowledgement}

\bibliography{references}

\includepdf[pages=-]{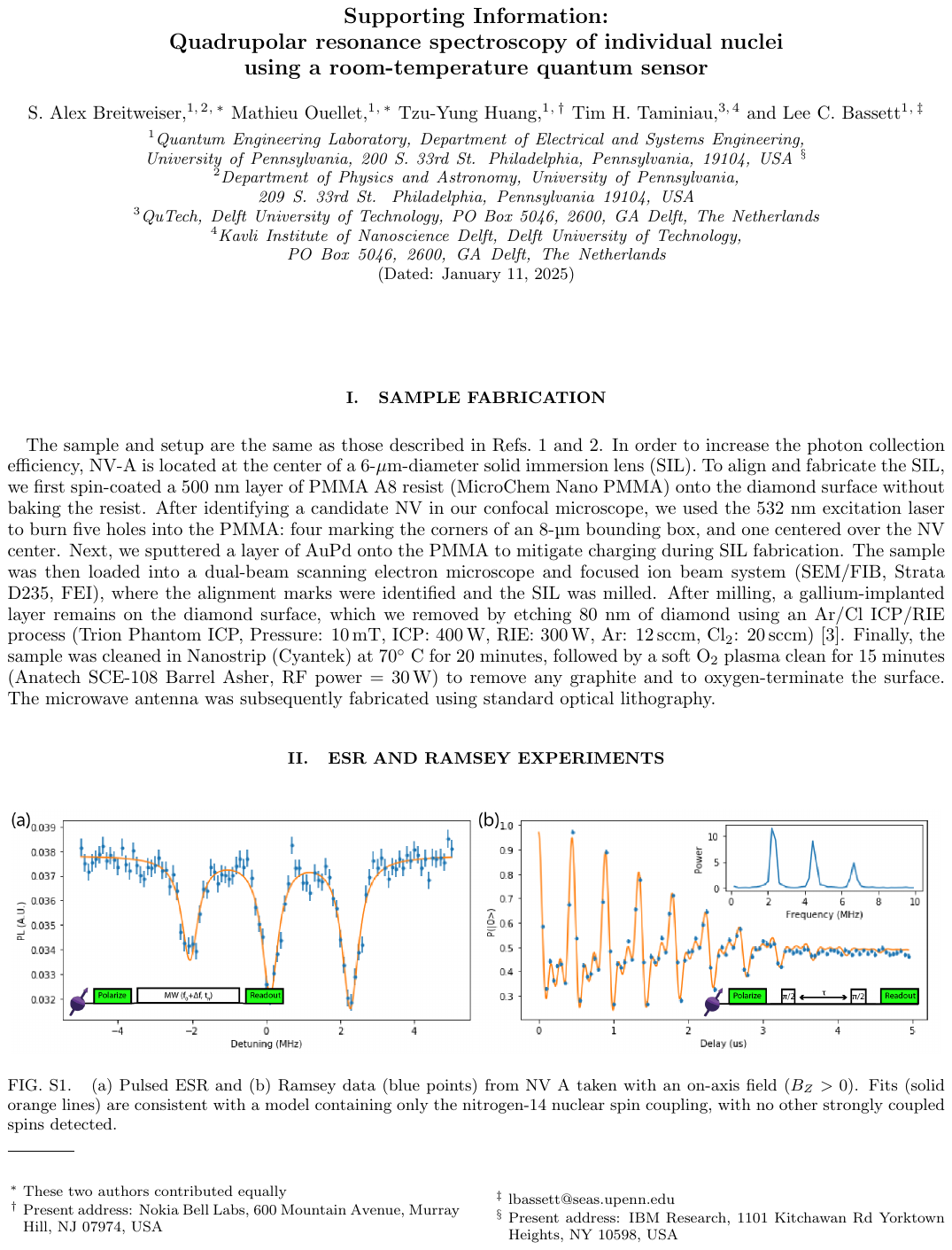}

\begin{figure}
\includegraphics[width=0.95\linewidth]{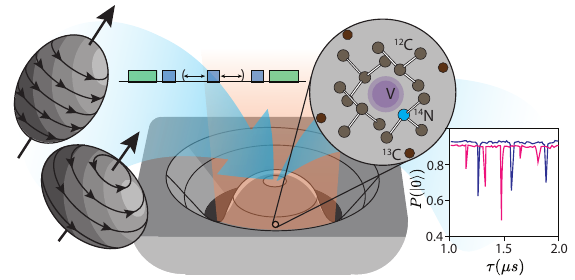}
\caption{\label{toc} 
\textbf{TOC Graphic} } 
\end{figure}

\end{document}